\pdfoutput=1

\documentclass[aps,pra,10pt,twocolumn,showpacs,superscriptaddress]{revtex4}
\usepackage{amsmath}
\usepackage{latexsym}
\usepackage{amssymb}
\usepackage{graphicx}
\usepackage[colorlinks=true]{hyperref}
\usepackage{float}

\begin{document}
\title{Lack of measurement independence can simulate quantum correlation even when signaling cannot}

\author{Manik Banik}
\email{manik11ju@gmail.com}
\affiliation{Physics and Applied Mathematics Unit, Indian Statistical Institute, 203 B.T. Road, Kolkata-700108, India}

\begin{abstract}
In Bell scenario, any nonlocal correlation, shared between two spatially separated parties, can be modeled deterministically either by allowing communications between the two parties or by restricting their free will in choosing the measurement settings. Recently, Bell scenario has been generalized into `semi-quantum' scenario where external quantum inputs are provided to the parties. We show that in `semi-quantum' scenario, entangled states produce correlations whose deterministic explanation is possible only if measurement independence is reduced. Thus in simulating quantum correlation `semi-quantum' scenario reveals a qualitative distinction between signaling and measurement dependence which is absent in Bell scenario. We further show that such distinction is not observed in `steering game' scenario, a special case of `semi-quantum' scenario. 
\end{abstract}


\maketitle
\section{Introduction}
In the early days of quantum mechanics, Einstein and his colleagues designed a thought experiment \cite{Einstein} intended to reveal what they believed to be inadequacies of quantum mechanics and they dreamed a causal world where no intrinsic randomness would present. But J.S. Bell showed that if experimenters enjoy complete free will in choosing measurement settings then quantum mechanics is indeed in contradiction with \emph{local realism} \cite{Bell}. Later in \cite{Jarrett}, it has been shown that Bell's locality condition can be thought of as a conjunction of two independent conditions namely outcome independence and parameter independence, popularly called no signaling, which restricts instantaneous communication between two spatially separated locations. In recent times, lots of interest have been found in simulating nonlocal correlations by violating these assumptions individually or collectively \cite{Hall_1,Hall_3,Kar,Barrett}; and in particular for singlet state correlation there exists several interesting simulation protocols \cite{Massar,Toner,Degorre,Cerf,Hall_2,Banik}. Please note, in Bell scenario any nonlocal correlation can be simulated by violating only the no signaling assumption (i.e. allowing communication between two parties). Moreover, for every such nonlocal correlation, there exists a different simulation protocol by violating only measurement independence condition. Therefore in simulating Bell nonlocal correlations there is no \emph{qualitative} difference between these two simulation resources. Interestingly, in this paper we show that this is not true in a more general situation called `semi-quantum' scenario which has been recently introduced in \cite{Buscemi}. We find that in semi-quantum scenario entangled quantum states reproduce correlations which can be simulated by reducing measurement independence but not possible by allowing communication. Further we prove that such qualitative distinction between no signaling and measurement independence is absent in `steering game' scenario \cite{Cavalcanti}, a particular case of semi- quantum scenario.     

\begin{figure}[h!]
\centering
\includegraphics[height=8cm,width=8cm]{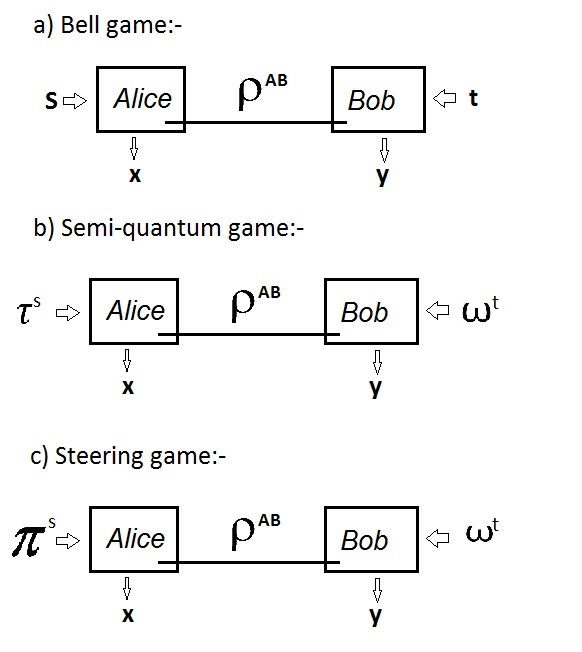}
\caption{a) In \emph{Bell game}, referee asks `classical
    questions' $s\in\mathcal{S}$ and $t\in\mathcal{T}$ to Alice and Bob respectively; which can be modeled
    by sending orthogonal states $\{\pi^s\}_{s\in\mathcal{S}}$ and $\{\pi^t\}_{t\in\mathcal{T}}$
    to Alice and Bob respectively. b) In \emph{semi-quantum game}, quantum states $\{\tau^s\}_{s\in\mathcal{S}}$ and $\{\omega^t\}_{t\in\mathcal{T}}$ (in general non-orthogonal) are sent to Alice and Bob respectively. c) In \emph{steering game}, Alice is given orthogonal states $\{\pi^s\}_{s\in\mathcal{S}}$ whereas Bob is given non-orthogonal states $\{\omega^t\}_{t\in\mathcal{T}}$.}\label{fig1}
\end{figure}

The Bell game scenario \cite{Bell,Clauser} involves two spatially separated players, say Alice and Bob. A referee picks indices $s\in\mathcal{S}$ and $t\in\mathcal{T}$ at random and sends them separately to Alice and Bob, respectively. The two players must compute answers $x\in\mathcal{X}$ and $y\in\mathcal{Y}$, respectively, and send the results to the referee, who will then pay them both. First, the players are told the rules of the game. Knowing the rules, the players are allowed to agree on any strategy and to share any variable $\lambda\in\Lambda$. The distribution $\rho(\lambda)$ of the shared variable $\lambda$ must be independent of the referee's questions, say $s$ and $t$; otherwise the measurement independence assumption will be violated. After starting the game the two players are forbidden to communicate with each other. Recently, F. Buscemi has generalized the Bell game into `semi-quantum' game \cite{Buscemi}. In this generalized scenario, instead of classical indices $s$ and $t$ referee sends quantum states $\tau^s\in\mathcal{H}_{A'}$ and $\omega^t\in\mathcal{H}_{B'}$ to Alice and Bob, respectively, without revealing the classical indices $s$ and $t$. Here $\mathcal{H}_{A'}$ and $\mathcal{H}_{B'}$ be the Hilbert spaces from where referee chooses input states for Alice and Bob, respectively. In general the set of quantum states $\{\tau^s\}_{s\in\mathcal{S}}$ and $\{\omega^t\}_{t\in\mathcal{T}}$ contain non orthogonal states. Whenever referee chooses quantum states from a set of orthogonal states the situation turns out to be the Bell game scenario. In between semi-quantum scenario and Bell scenario another interesting situation has been considered in \cite{Cavalcanti}, namely `steering game' scenario. In steering game referee sends orthogonal quantum state $\pi^s$ to one party (say Alice) and send quantum states $\omega^t$ (in general non orthogonal) to other party (see Fig.\ref{fig1}).  

Recently, Rosset \emph{et.al.} proved a strong no-go result regarding simulation of correlations achieved in semi-quantum scenario \cite{Rosset}. They have shown that any entangled state can generate correlations in such simulation task which cannot be simulated by local operation assisted by classical communication (LOCC) \cite{Bennett1,Bennett2} even if there is no limitation on classical communication. Such a strong no-go result is possible in semi-quantum scenario, as referee sends nonorthogonal quantum states to Alice and Bob, who will not be able to discriminate the states even if they are allowed to communicate with each other. Interestingly, we show that a simulation protocol exists for such correlation if measurement independence assumption is reduced by suitable amount.  

\section{simulation task in semi-quantum scenario}
In semi-quantum scenario the simulation task is defined via the 5-tuple $(\varrho_{AB}, \{\tau^s_{A'}\},\{\omega^t_{B'}\}, \{\mathcal{A}^x_{A'A}\}, \{\mathcal{B}^y_{BB'}\})$; where $\varrho_{AB}$ is the shared entangled state between the two players Alice and Bob, referee sends quantum inputs $\{\tau^s_{A'}\}$ and $\{\omega^t_{B'}\}$ to Alice and Bob respectively, $\{\mathcal{A}^x_{A'A}\}$ and $\{\mathcal{B}^y_{BB'}\}$ are positive-operator-valued-measure (POVM) \cite{Nielsen} elements acting respectively on the composite quantum systems $A'A$ and $BB'$. Consider the state $\varrho_{AB}$ of dimension $d_A\times d_B$, and Alice and Bob perform local measurements $\mathcal{A}^x_{A'A}$ and $\mathcal{B}^y_{BB'}$ with binary outcomes $x,y\in\{1,0\}$; where the outcome $x=1$ ($y = 1$) corresponds to the successful projection onto the maximally entangled state $|\phi^{+}_d\rangle=\Sigma^{d-1}_{k=0}|kk\rangle/\sqrt{d}$, i.e.
\begin{eqnarray}
\mathcal{A}^1_{A'A}=|\phi^{+}_{d_A}\rangle\langle\phi^{+}_{d_A}|~,~~~~~\mathcal{A}^{0}_{A'A}=\mathbb{I}-|\phi^{+}_{d_A}\rangle\langle\phi^{+}_{d_A}|;\\
\mathcal{B}^1_{BB'}=|\phi^{+}_{d_B}\rangle\langle\phi^{+}_{d_B}|~,~~~
~~\mathcal{B}^{0}_{BB'}=\mathbb{I}-|\phi^{+}_{d_B}\rangle\langle\phi^{+}_{d_B}|.
\end{eqnarray}
The input states are chosen to have the same dimensions as the respective subsystem of A and B in $\varrho_{AB}$, $|\tau^s\rangle\in \mathbb{C}^{d_A}$, $|\omega^t\rangle\in\mathbb{C}^{d_B}$, and constructed such that the corresponding density matrices $\{|\tau^s\rangle\langle\tau^s|\}$, $\{|\omega^t\rangle\langle\omega^t|\}$ span the space of linear operators acting on $\mathbb{C}^{d_A}$ and $\mathbb{C}^{d_B}$ respectively. For more specific example of semi-quantum game let Alice and Bob share singlet states $\varrho_{AB}=|\psi^-\rangle\langle\psi^-|$ and consider that the referee chooses the quantum states from the vertices of a regular tetrahedron on the surface of the Bloch sphere. Using $\vec{\sigma}=(\sigma_1,\sigma_2,\sigma_3)$ the vector of Pauli matrices, and $\vec{v_1}=\frac{(1,1,1)}{\sqrt(3)}$, $\vec{v_2}=\frac{(1,-1,-1)}{\sqrt(3)}$, $\vec{v_3}=\frac{(-1,1,-1)}{\sqrt(3)}$, $\vec{v_4}=\frac{(-1,-1,1)}{\sqrt(3)}$, we denote, for $s,t=1,2,3,4$:
\begin{equation}
|\tau^s\rangle\langle\tau^s|=\frac{\mathbb{I}+\vec{v_s}.\vec{\sigma}}{2}~,~
~~|\omega^t\rangle\langle\omega^t|=\frac{\mathbb{I}+\vec{v_t}.\vec{\sigma}}{2}.
\end{equation}
The resulting correlations are then:
\begin{eqnarray}\label{qicorre}
P_{|\psi^-\rangle}(0,0||\tau^s\rangle,|\omega^t\rangle)=\left\{\begin{array}{lll}~~\frac{1}{2}~~\mbox{if s=t}\\~~\\~~\frac{7}{12}~~\mbox{otherwise.}
\end{array}\right.\nonumber\\
P_{|\psi^-\rangle}(0,1||\tau^s\rangle,|\omega^t\rangle)=\left\{\begin{array}{lll}~~\frac{1}{4}~~\mbox{if s=t}\\~~\\~~\frac{1}{6}~~\mbox{otherwise.}
\end{array}\right.\nonumber\\
P_{|\psi^-\rangle}(1,0||\tau^s\rangle,|\omega^t\rangle)=\left\{\begin{array}{lll}~~\frac{1}{4}~~\mbox{if s=t}\\~~\\~~\frac{1}{6}~~\mbox{otherwise.}
\end{array}\right.\nonumber\\
P_{|\psi^-\rangle}(1,1||\tau^s\rangle,|\omega^t\rangle)=\left\{\begin{array}{lll}~~0~~\mbox{if s=t}\\~~\\~~\frac{1}{12}~~\mbox{otherwise.}
\end{array}\right.
\end{eqnarray}
The referee randomly chooses quantum states and sends to Alice and Bob who can share classical variable $\Lambda$ but the distribution of the classical variable does not depend on referee's input. As the input states are nonorthogonal Alice and Bob cannot perfectly discriminate the states. Moreover communications between Alice and Bob, even in infinite amount, have no use in discriminating those nonorthogonal input states. As a result no LOCC simulation protocol is possible for the correlation (\ref{qicorre}) \cite{Rosset}. Interestingly, in the following section we show that to have a simulation protocol for the correlation (\ref{qicorre}) Alice and Bob need not to know the referee's quantum input at all. Rather using the correlation in the input states they can easily simulate the correlation-(\ref{qicorre}) if the distribution of the variable $\Lambda$ is made dependent on referee's inputs i.e. violating the measurement independence assumption. 

\section{reduced measurement independence model}
Let Alice and Bob share hidden variable $\Lambda$, taking $4$ distinct values $\lambda=\lambda_1,\lambda_2,\lambda_3,\lambda_4\in\Lambda$.  These variables are distributed according to one of two probability distributions $p_{\_Sa(\lambda)}$ and $p_{\_Di(\lambda)}$ :\\
for $s=t$
\begin{equation}\label{distri_1}
\{p_{\_Sa(\lambda)}\}\equiv\{p(\lambda|s,t)\}:=\{1/2,1/4,1/4,0\}
\end{equation}
and for $s\neq t$
\begin{equation}\label{distri_2}
\{p_{\_Di(\lambda)}\}\equiv\{p(\lambda|s,t)\}:=\{7/12,1/6,1/6,1/12\}
\end{equation}
The outcomes of Alice and Bob are fully determined by the parameter $\Lambda$ 
\begin{equation}
x=f(\lambda),~~~~y=g(\lambda)
\end{equation}
where,
\begin{eqnarray*}
f(\lambda_1)=f(\lambda_2)=0,~~f(\lambda_3)=f(\lambda_4)=1,\\
g(\lambda_1)=g(\lambda_3)=0,~~g(\lambda_2)=g(\lambda_4)=1.
\end{eqnarray*}
The corresponding correlations are of the form
\begin{eqnarray}\label{simu}
 p(x,y|s,t) &=& \sum_{\lambda} p(x,y|x,t,\lambda) p(\lambda|s,t)\nonumber\\
 &=& \sum_{\lambda}  \delta_{x,f(\lambda)}  \delta_{y,g(\lambda)} p(\lambda|s,t)
\end{eqnarray}
where $\delta_{0,0}=\delta_{1,1}=1$ and $\delta_{0,1}=\delta_{1,0}=0$. It is clear that correlation (\ref{simu}) is same as correlation (\ref{qicorre}). 

The lack of measurement independence required to reproduce the correlation (\ref{qicorre}) can be quantified in several ways. Measurement independence is the property that the distribution of the underlying variable is independent of the measurement settings or referee's input question, i.e.,
\begin{equation}
p(\lambda|s,t)=p(\lambda|s',t')
\end{equation}
Bays theorem guarantees that is condition is equivalent to $p(s,t|\lambda)=p(s,t)$, which is often justified by the notion of experimental free will, i.e., that experimenters can freely choose between different measurement settings irrespective of the underlying variable $\lambda$ describing the system. The degree of measurement dependence is quantified by the variational distance \cite{Hall_1,Hall_2}
\begin{equation}
M :=\sup_{s,s',t,t'}\sum_{\lambda\in\Lambda}|p(\lambda|s,t)-p(\lambda|s',t')|.
\end{equation}
Clearly $0\leq M\leq 2$, where two extreme values signify complete measurement independence and complete measurement dependence, respectively. The fraction of measurement independence or free will is defined by
\begin{equation}
F :=1-M/2.
\end{equation}
Thus, $0\leq F\leq 1$, with $F = 0$ corresponding the case where no experimental free will can be enjoined and $F = 1$ corresponding the case with complete experimental free will. For the above reduced free will model we have:
\begin{equation*}
M=|\frac{7}{12}-\frac{1}{2}|+2|\frac{1}{6}-\frac{1}{4}|+|\frac{1}{12}-0|=\frac{1}{3},
\end{equation*}
and the amount of experimental free will is given by
\begin{equation*}
F=1-\frac{1}{2.3}\cong 83.3\%.
\end{equation*}
The degree of measurement dependence of the model can also be quantified by the mutual information between the labels sent by the referee and the shared variable $\Lambda$ which is defined as \cite{Hall_1,Barrett}:
\begin{equation}\label{mu}
H(\mathcal{S},\mathcal{T}:\Lambda) = \sum_{s,t,\lambda} p(s,t,\lambda)\log_2 \frac{p(s,t,\lambda)}{p(s,t)p(\lambda)}.
\end{equation}
Maximizing the mutual information over all possible distributions of measurement settings the measurement dependence capacity \cite{Hall_1} of a given model can be obtained i.e.
\begin{equation}
\mathcal{C}_{\mbox{meas dep}}:=\sup_{p(s,t)}H(\mathcal{S},\mathcal{T}:\Lambda).
\end{equation}
Denote $P (= \sum_{s=t} p(s,t))$ as the probability of sending the same label to Alice and Bob. Then for our model the  Eq.(\ref{mu}) becomes
\begin{eqnarray}
H(\mathcal{S},\mathcal{T}:\Lambda)= H[ P p_{\_Sa(\lambda)} + (1-P) p_{\_Di(\lambda)} ]\nonumber\\
- P H[p_{\_Sa(\lambda)}] - (1-P) H[p_{\_Di(\lambda)}].
\end{eqnarray}
\begin{figure}[h!]
\centering
\includegraphics[height=4cm,width=6cm]{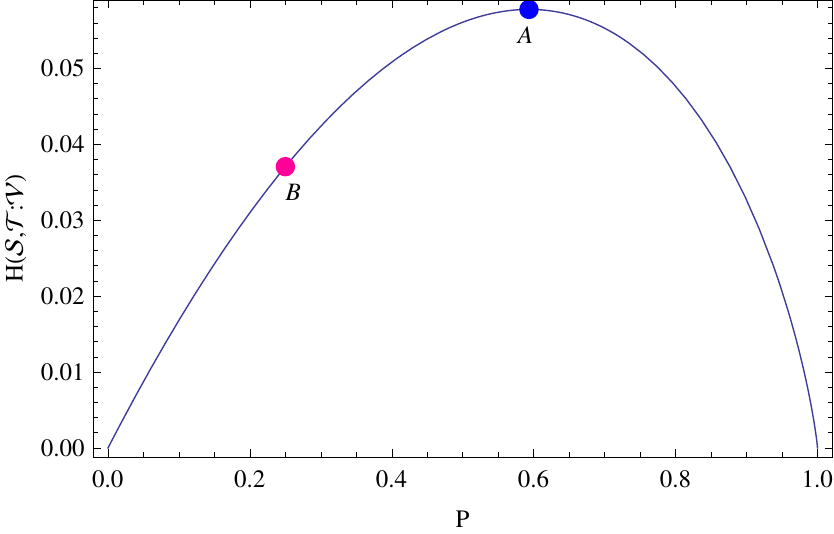}
\caption{$H(\mathcal{S},\mathcal{T}:\Lambda)$ Vs $P$ plot. The point $A\cong(0.593,0.057)$ correspond to maximum
mutual information i.e., \emph{measurement dependence capacity} of the model; and the point $B\cong(0.25,0.037)$
correspond to mutual information between the variable $\Lambda$ and $(\mathcal{S},\mathcal{T})$ when $p(s,t)$ is uniform.}\label{fig}
\end{figure}
From Fig.\ref{fig} it is clear that measurement dependence capacity of our model turns out to be $ 0.05778$ whereas for uniform distributions of measurement settings ($p(s,t)=\frac{1}{16}$ or $P=\frac{1}{4}$) the mutual information becomes $0.03705$.

Please note, distribution of the variable $\Lambda$ depends on referee's input in such a way (Eq.(\ref{distri_1}) and Eq.(\ref{distri_2})) that having accesses the variable $\Lambda$ neither Alice nor Bob can guess the classical indices of their quantum states. Actually $\Lambda$ does not contain any information about the individual quantum state send to Alice or Bob; it only contains information about the correlation in the input states send to Alice and Bob and this is sufficient to provide the above simulation protocol. Thus for correlation-(\ref{qicorre}) communication based simulation protocol is impossible while we provide a simulation protocol by sacrificing measurement independence. Thus simulating the above quantum correlations (correlation-(\ref{qicorre})) communication has no use while lack of measurement independence becomes useful---and hence it proves a qualitative distinction between no signaling and measurement independence (or free will). Our simulation model also makes it clear that the no-go result obtained in \cite{Rosset} have been derived under free will assumption, which assumes no correlation between the referee's labels $(s,t)$ and any other physical parameter $\Lambda$ that propagates to the future (and hence to Alice and Bob). This is a very strong assumption, and our result further shows that even a small amount of lack of measurement independence allows simulation of the correlations by non-quantum means without any need of communication.

\section{locc simulation of steering game correlations}
Other than non-locality, a much discussed non classical feature of quantum mechanics is steering, first introduced by Schr\"{o}dinger \cite{Schro} and recently brought in attention by Wiseman and his collaborators \cite{Wiseman}. In \cite{Cavalcanti}, Cavalcanti \emph{et al.} introduce the `steering game' as one of the interesting special case of semi-quantum game. In steering game the referee sends the orthogonal quantum states $\pi^s\in\mathcal{H}_{A'}$ to one party (say Alice) and sends quantum states $\omega^t\in\mathcal{H}_{B'}$ (in general non orthogonal) to Bob, without revealing the classical indices $s\in\mathcal{S}$ and $t\in\mathcal{T}$. In the context of our work it is an interesting question whether in steering game scenario qualitative distinction between measurement independence and no signaling is possible. We answer this question in negation by providing a LOCC simulation protocol for all such steering games. 

Given the bipartite state $\rho_{AB}$ shared  between Alice and Bob, let us denote the set of all joint quantum conditional probabilities by $\mu(\rho_{AB}):=\{\mu(x,y|s,t)~| ~x\in\mathcal{X};~y\in\mathcal{Y};~s\in\mathcal{S};~t\in\mathcal{T}\}$, where $x\in\mathcal{X}$ and $y\in\mathcal{Y}$ are classical results of Alice and Bob respectively that has to convey to the referee. $\mu(x,y|s,t)$ is computed as:
\begin{equation*}
\mu(x,y|s,t)=Tr[(P_{A_0A}^x\otimes
      Q_{BB_0}^y)(\pi^s_{A_0}\otimes\rho_{AB}\otimes\omega^t_{B_0})],
\end{equation*}
where $P\equiv(P^x_{A_0A};x\in\mathcal{X}~|~\sum_{x\in\mathcal{X}}P^x_{A_0A}=\textbf{1})\in\mathcal{M}(A_0A;\mathcal{X})$ and $Q\equiv(Q^y_{B_0B};y\in\mathcal{Y}~|~\sum_{y\in\mathcal{Y}}Q^y_{B_0B}=\textbf{1})\in\mathcal{M}(BB_0;\mathcal{Y})$. By $\mathcal{M}(A;\mathcal{X})$, we denote the convex set of all $\mathcal{X}$-POVMs on $A$.

By simulation of the correlations $\mu(\rho_{AB})$, we mean that Alice and Bob are not allowed to share the state $\rho_{AB}$, but they have to reproduce the correlations in a given operational paradigm like LOSR or LOSSR or LOCC. Now if the state $\rho_{AB}$ is steerable then according to the result of Cavalcanti \emph{et.al} \cite{Cavalcanti} we know that $\mu(\rho_{AB})$ cannot be simulated by \emph{local operation with steering and shared randomness} (LOSSR) resources and therefore by LOSR resources. Interestingly if there allowed operational paradigm is LOCC then they will be succeed in the simulation task.

\subsection*{LOCC simulation protocol}
\begin{enumerate}
\item[(I)] Bob locally prepares the state $\rho_{AB}$ in his laboratory and takes another particle described by the Hilbert space $\mathcal{H}_{B_0}\cong\mathcal{H}_{A'}$.
\item[(II)] Receiving orthogonal quantum states $\pi^s$'s from referee, Alice reveals the classical induces $s$ by performing Von-Newman measurement in $\{\pi^s\}_{s\in\mathcal{S}}$ basis.
\item[(III)] Alice communicates her induces to Bob by using $\log_2\mathfrak{C(\mathcal{S})}$ bits, where $\mathfrak{C(\mathcal{S})}$ be the cardinality of the set $\mathcal{S}$.
\item[(IV)] Receiving the information of Alice's index $s$, Bob prepares the system $B_0$ in the state $\pi^s$.
\item[(V)]  Receiving the quantum states $\omega^t$'s from referee, Bob performs the required POVM $P\in\mathcal{M}(AB_0;\mathcal{X})$ and $Q\in\mathcal{M}(BB';\mathcal{Y})$ and reproduces the joint conditional probability distributions $p(x,y|s,t)$.
\item[(VI)] Bob communicates the result $x$ to Alice by using $\log_2\mathfrak{C(\mathcal{X})}$ bits.
\end{enumerate}  
Note that, to perform the simulation task successfully, Alice and Bob require both way classical communications. If the simulation task demands that Alice and Bob have to reproduce the correlation only and does not further demand that the outcomes $x\in\mathcal{X}$ and $y\in\mathcal{Y}$ have to be supplied to referee by the respective parties, then Bob needs not communicate the result $x$ to Alice. He sends both of the results $x$ and $y$ to referee himself and in this case one way communication ($\log_2\mathfrak{C(\mathcal{S})}$ bits from Alice to Bob) suffices to complete the simulation task. To perform the simulation task, Bob needs to have the resources in order to prepare the suitable bipartite state $\rho_{AB}$ that exhibits steering. But this might be particularly relevant in a scenario where Alice and Bob want to pretend to be able to share such a state over long distances. In this scenario, they can reproduce the correlation without sharing the state and Bob keeping it with him. Actually by the above clever means, they will be able to preserve sufficient entanglement shared between them. Thus following the above protocol, correlations achieved in all steering games can be simulated in LOCC operational framework with bounded amount of classical communication. One can also simulate such correlations without using communication but making the distribution of the classical variable $\Lambda$ dependent on the referee's input. Thus in simulating correlations, achieved in steering game scenario, signaling and lack of measurement independence have same qualitative footing.

\section{conclusion}
Experimental free will and no-signaling are two very important physically motivated assumptions. Various important no-go results in quantum foundation have been derived under these assumptions. While Bell's theorem \cite{Bell} and recently proved Colebeck and Renner's result \cite{Colbeck} consider both the assumptions, Kochen-Specker no-go result \cite{Kochen} uses only free will assumption. In an interesting discussion \cite{Bub} of the measurement problem for a PR-box, J. Bub pointed out that, if Bob has access to a parameter or ontic state $\lambda$ which completely specifies the output values for both of his possible inputs, or even specifies whether these output values are the same or different, then either there is a violation of no-signaling principle or Alice's choice of input is not free but depends on the value of $\lambda$. Actually Bub's conclusion holds for any Bell non-local correlation. In this paper we provide a simulation protocol with relaxed measurement independence for correlation achieved in a semi-quantum game whereas no simulation protocol involving only communication but no measurement dependence is possible for this correlations. Thus our result proves that in `semi-quantum' scenario there exists correlation whose deterministic explanation is possible only if measurement independence assumption is relaxed and hence it establishes a qualitative distinction between no signaling and measurement independence. We further show that such distinction is not possible in steering game scenario. To prove the optimality of our model is an interesting open question. One may also get interested in constructing optimal reduced free will model for other semi-quantum games.   

{\bf Acknowledgment}: The author acknowledges useful discussions with A. Rai, S. Das, Md R. Gazi, S. Kunkri and is grateful to G. Kar for discussion about the LOCC simulation protocol for quantum refereed steering correlations, and also gratefully acknowledges Michael J.W. Hall for suggesting a simple version of the reduced free will model. It is also my pleasure to thank F. Buscemi for useful discussion about his work \cite{Buscemi}.


\begin{thebibliography}{99}
\bibitem{Einstein} A. Einstein, B. Podolsky, and N. Rosen,Phys. Rev. {\bf 47}, 777 (1935).
\bibitem{Bell} J. S. Bell, Physics {\bf 1}, 195 (1964).
\bibitem{Jarrett} J. P. Jarrett, Nous {\bf 18(4)}, 569 (1984).
\bibitem{Hall_1} M.J.W. Hall, Phys. Rev. A {\bf 84}, 022102 (2011).
\bibitem{Hall_3} M.J.W. Hall,  Phys. Rev. A {\bf 82}, 062117 (2010).
\bibitem{Barrett} J. Barrett, N. Gisin, Phys. Rev. Lett. {\bf 106}, 100406, (2011).
\bibitem{Kar} G. Kar, MD. R. Gazi, M. Banik, S. Das, A. Rai, S. Kunkri,  J. Phys. A: Math. Theor. {\bf 44}, 152002 (2011).
\bibitem{Massar} S. Massar, D. Bacon, N. Cerf, and R. Cleve, Phys. Rev.A {\bf 63}, 052305 (2001).
\bibitem{Toner} B.F. Toner and D. Bacon, Phys. Rev. Lett. {\bf 91}, 187904 (2003).
\bibitem{Degorre} J. Degorre, S. Laplante, and J. Roland, Phys. Rev. A {\bf 72}, 062314 (2005).
\bibitem{Cerf} N.J. Cerf, N. Gisin, S. Massar, and S. Popescu, Phys. Rev. Lett. {\bf 94}, 220403 (2005).
\bibitem{Hall_2} M.J.W. Hall, Phys. Rev. Lett. {\bf 105}, 250404 (2010).
\bibitem{Banik} M. Banik, MD R. Gazi, S. Das, A. Rai, S. Kunkri, J. Phys. A: Math. Theor. {\bf 45}, 205301 (2012).
\bibitem{Buscemi} F. Buscemi, Phys. Rev. Lett. {\bf 108}, 200401 (2012).
\bibitem{Cavalcanti} E. G. Cavalcanti, M. J. W. Hall, and H. M. Wiseman, Phys. Rev. A {\bf 87}, 032306 (2013).
\bibitem{Clauser} J. F. Clauser, M.A. Horne, A. Shimony and R. A. Holt, Phys. Rev. Lett. {\bf 23}, 880 (1969).
\bibitem{Rosset} D. Rosset, C. Branciard, N. Gisin, and Y. C. Liang, New J. Phys. {\bf 15} 053025 (2013).
\bibitem{Bennett1} C. H. Bennett, G. Brassard, S. Popescu, B. Schumacher, J. A. Smolin, and W. K. Wootters, Phys. Rev. Lett. {\bf 76}, 722 (1996);
\bibitem{Bennett2} C. H. Bennett, D. P. DiVincenzo, J. A. Smolin, and W. K Wootters, Phys. Rev. A {\bf 54}, 3824 (1996).
\bibitem{Nielsen} M. A. Nielsen and I. L. Chuang, Quantum Computation and Quantum Information (Cambridge Univ. Press, New York, 2000).
\bibitem{Schro} E. Schrodinger, Proc. Cambridge Philos. Soc. {\bf 31}, 553 (1935); {\bf 32}, 446 (1936).
\bibitem{Wiseman} H. M. Wiseman, S. J. Jones and A. C. Doherty, Phys. Rev. Lett. {\bf 98}, 140402 (2007); S. J. Jones, H. M. Wiseman, and A. C. Doherty, Phys. Rev. A {\bf 76}, 052116 (2007); E. G. Cavalcanti, S. J. Jones, H. M. Wiseman, M. D. Reid, Phys. Rev. A {\bf 80}, 032112 (2009).
\bibitem{Colbeck} R. Colbeck, R. Renner, Nature Communications {\bf 2}, 411 (2011);  R. Colbeck, R. Renner, Phys. Rev. Lett. {\bf 108}, 150402 (2012).
\bibitem{Kochen} S. Kochen and E. P. Specker, J. Math. Mech. {\bf 17}, 59 (1967); J. Bub, Interpreting the Quantum World (Cambridge University Press, 1999); A. Peres, J. Phys. A: Math. Gen. {\bf 24}, L175 (1991); N.D. Mermin, Phys. Rev. Lett. {\bf 65}, 3373 (1990); A. Cabello, Nature {\bf 474}, 456 (2011).
\bibitem{Bub} J. Bub,  arXiv [quant-ph]:1210.6371.
\end{thebibliography}
\end{document}